\documentclass[
reprint,
superscriptaddress,
amsmath,amssymb,
aps,
pre,
]{revtex4-2}

\usepackage{graphicx}
\usepackage{dcolumn}
\usepackage{bm}
\usepackage{subcaption}
\usepackage{float}
\usepackage{xcolor}
\usepackage[justification=raggedright, singlelinecheck=false]{caption}
\usepackage{comment}
\usepackage{hyperref}
\usepackage{cleveref}
\usepackage{booktabs}
\usepackage{comment}

\hypersetup{colorlinks=True, linkcolor=gray, citecolor=gray, urlcolor=gray}
\crefformat{equation}{Eq.~(#2#1#3)}

\makeatletter
\renewcommand\@fnsymbol[1]{%
  \ifcase#1\or
    \normalfont e\relax
  \else
    \@ctrerr
  \fi
}
\makeatother

\newcommand{\ra}[1]{\renewcommand{\arraystretch}{#1}}

\newcommand{\rev}[1]{#1}

\begin{document}

\title{Learned Free-Energy Functionals from Pair-Correlation Matching\\for Dynamical Density Functional Theory}

\author{Karnik Ram}
\email{karnik.ram@tum.de}
\affiliation{School of Computation, Information and Technology, TU Munich, Germany}
\affiliation{Munich Center for Machine Learning, Germany}
\author{Jacobus Dijkman}
\affiliation{Van't Hoff Institute for Molecular Sciences, University of Amsterdam, the Netherlands}
\affiliation{Informatics Institute, University of Amsterdam, the Netherlands}
\author{René van Roij}
\affiliation{Institute for Theoretical Physics, Utrecht University, the Netherlands}
\author{Jan-Willem van de Meent}
\affiliation{Informatics Institute, University of Amsterdam, the Netherlands}
\author{Bernd Ensing}
\affiliation{Van't Hoff Institute for Molecular Sciences, University of Amsterdam, the Netherlands}
\author{Max Welling}
\affiliation{Informatics Institute, University of Amsterdam, the Netherlands}
\author{Daniel Cremers}
\affiliation{School of Computation, Information and Technology, TU Munich, Germany}
\affiliation{Munich Center for Machine Learning, Germany}

\date{\today}

\begin{abstract}
Classical density functional theory (cDFT) and dynamical density functional theory (DDFT) are modern statistical mechanical theories for modeling many-body colloidal systems at the one-body density level. The theories hinge on knowing the excess free-energy accurately, which is however not feasible for most practical applications. Dijkman \textit{et al.}~\href{https://doi.org/10.1103/PhysRevLett.134.056103}{[Phys. Rev. Lett. 134, 056103 (2025)]} recently showed how a neural excess free-energy functional for cDFT can be learned from bulk simulations via pair-correlation matching. In this article, we demonstrate how this same functional can be applied to DDFT, without any retraining, to simulate non-equilibrium overdamped dynamics of inhomogeneous densities. We evaluate this on a three-dimensional Lennard-Jones system with planar geometry under various complex external potentials and observe good agreement of the dynamical densities with those from expensive Brownian dynamic simulations, up to the limit of the adiabatic approximation. We further develop and apply an extension of DDFT based on gradient flows, to a grand-canonical system modeled after breakthrough gas adsorption studies, finding similarly good agreement. Our results demonstrate a practical route for leveraging learned free-energy functionals in DDFT, paving the way for accurate and efficient modeling of many-body non-equilibrium systems.
\end{abstract}
\maketitle

\section{Introduction}
Studying the collective behavior of many-body systems emerging from inter-particle interactions is fundamental in many diverse fields of science, ranging from gas adsorption studies in porous media~\cite{Soares2023} to semiconductors and protein adsorption~\cite{wei2014protein}. Classical density functional theory (cDFT)~\cite{evans} is a promising theory for accurately and efficiently modeling such many-body systems at the level of the one-body density profile $\rho(\mathbf{r})$ in thermodynamic equilibrium. cDFT relies on the existence of an excess free-energy functional $\mathcal{F}_\text{exc}[\rho(\mathbf{r})]$ which describes the inter-particle interactions, accounting for the non-ideal contribution to the total Helmholtz intrinsic free-energy functional $\mathcal{F}[\rho(\mathbf{r})]$. This functional is generally unknown and various approximations have been constructed ranging from the mean-field approximation~\cite{hansen} for soft interactions to fundamental measure theory~\cite{fmt} for hard-sphere interactions. More recently, neural functionals learned from Monte Carlo simulations have been proposed~\cite{linClassicalDensityFunctional2019, cats, florian, simonMachineLearningDensity2024a, dijdft} and exhibit remarkable accuracy.

Dynamical density functional theory (DDFT)~\cite{evans, marconi, archer2004dynamical, fundamentals} is an extension of cDFT to describe non-equilibrium systems as a Fokker-Planck equation on the one-body density profile. DDFT has been applied in diverse systems ranging from simple colloidal fluids~\cite{marconi} to plasmas~\cite{plasmas} and microswimmers~\cite{microswimmers}. A recent review of DDFT applications can be found in~\cite{perspective}. Like cDFT, DDFT also relies on the excess free-energy functional and typically the same equilibrium functional from cDFT is adopted for DDFT under the adiabatic approximation.

In this article, we adopt the neural functional proposed by Dijkman~\textit{et al.}~\cite{dijdft}, which was trained by matching the pair-correlation function from bulk simulations, and we evaluate it for DDFT. Despite only having seen bulk densities during training, we show that this functional produces accurate time evolution of $1$D inhomogeneous densities on a $3$D Lennard-Jones system with planar geometry. We further extend DDFT with a particle source term, based on gradient flows~\cite{ambrosio2008gradient, wasserstein}, to model grand-canonical systems with non-conserved dynamics. We evaluate this extension on a system inspired by gas adsorption simulations, and observe good agreement with a hybrid Monte Carlo Brownian dynamic simulation.

\begin{figure*}
    \includegraphics[width=0.9\textwidth]{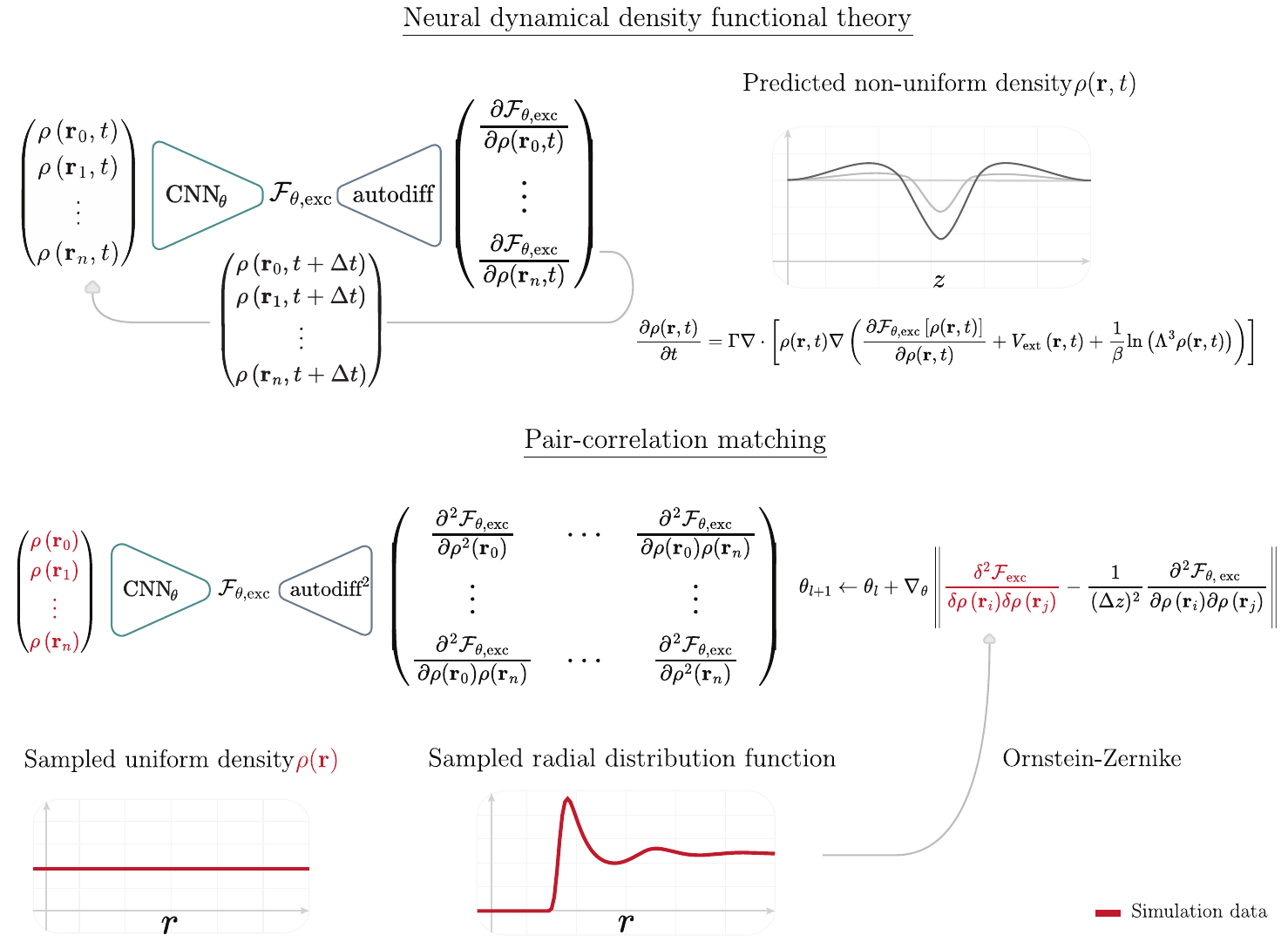}
    \caption{An overview of the proposed neural DDFT framework, and the pair-correlation matching~\cite{dijdft} approach. (Bottom) A neural free-energy functional is trained via pair-correlation matching where its second functional derivative is optimized against the direct pair correlation function computed from simulated uniform bulk densities. (Top) The neural excess free-energy functional $\mathcal{F}_{\theta, \text{exc}}$ is used for DDFT to time evolve the inhomogeneous particle density $\rho(\mathbf{r})$ of a 3D Lennard-Jones system with planar geometry via automatic differentiation and numerical integration, towards equilibrium.}
    \label{fig:block}
\end{figure*}

\section{Pair-correlation matching}\label{pc-matching}

Pair-correlation matching~\cite{dijdft} is a method for learning an accurate neural excess free-energy functional $\mathcal{F}_{\theta, \text{exc}}$ (where $\theta$ represents neural network parameters) by training exclusively on a dataset of bulk radial distribution functions, rather than from heterogeneous density profiles. The method uses the relationship between the second functional derivative of the excess free-energy functional  $\left(\delta^2 \mathcal{F}_{\text{exc}}/\delta \rho (z_i) \delta \rho (z_j)\right)$ and the direct correlation function $c^{(2)}(\mathbf{r})$, which is related to the radial distribution function $g(r)$ via the Ornstein-Zernike equation (\cite{hansen}, Chap.3), for training. Specifically, a neural network $\mathcal{F}_{\theta, \text{exc}}$ is trained by optimizing a physics-based regularizer~\cite{pinns}, which matches its Hessian $\left(\partial^2 \mathcal{F}_{\text{exc}}/\partial \rho_i \partial \rho_j\right)$ with the direct correlation function obtained from Monte Carlo simulations of homogeneous bulk systems (Fig.~\ref{fig:block}). This functional is then used in the cDFT scheme to obtain accurate estimates for inhomogeneous equilibrium density profiles under various external fields without having used any inhomogeneous densities during training.
In addition to pair-correlation matching, we can also train a neural free-energy functional using single-body direct correlation matching; this is further detailed with a brief recapitulation of the key cDFT equations in~\Cref{cdft_theory}.

\section{Classical dynamical density functional theory}

DDFT was first introduced phenomenologically by Evans~\cite{evans}, and later microscopically derived from the Langevin equations by Marconi and Tarazona ~\cite{marconi}, and from Smoluchowski equations by Archer and Evans~\cite{archer2004dynamical}. An extensive review of DDFT fundamentals can be found in~\cite{fundamentals}; see~\Cref{ddft-derivation} for a concise derivation. DDFT has the form of a continuity equation where $\partial_t \rho(\mathbf{r},t)$ is proportional to the divergence of a vector field as follows:

\begin{equation}\label{cddft}
    \frac{\partial \rho(\mathbf{r}, t)}{\partial t} = \Gamma \nabla \cdot\left[\rho(\mathbf{r}, t) \nabla\left(\frac{\delta \mathcal{F}[\rho]}{\delta \rho(\mathbf{r}, t)}\right)\right],
\end{equation} where $\Gamma$ is the mobility coefficient and $\mathcal{F}[\rho]$ is the intrinsic Helmholtz free-energy functional. This functional consists of three parts,
\begin{equation}
\begin{split}
\mathcal{F}[\rho] &= \frac{1}{\beta}\int d^3\mathbf{r}\rho(\mathbf{r})(\ln(\Lambda^3 \rho(\mathbf{r})) - 1) \\
&\qquad + \int d^3\mathbf{r} \rho(\mathbf{r}) V_{\text{ext}}(\mathbf{r}) + \mathcal{F}_{\text{exc}}([\rho]),
\end{split}
\end{equation}
where $\mathcal{F}_{\text{exc}}([\rho])$ is the excess free-energy functional, $V_{\text{ext}}\rev{(\mathbf{r})}$ is the external potential, $\beta = \left(1/k_BT\right)$, and $\Lambda$ is the thermal wavelength. Expanding~\cref{cddft} further we get,
\begin{align}\label{cddft_full}
    \frac{\partial \rho(\mathbf{r}, t)}{\partial t} &= \Gamma \nabla \cdot\Bigg[\rho(\mathbf{r}, t) \nabla \Bigg(\frac{\delta \mathcal{F}_\text{exc}[\rho]}{\delta \rho(\mathbf{r}, t)}
   + V_{\text{ext}}(\mathbf{r},t)  \nonumber \\
   &\qquad\qquad + \frac{1}{\beta} \ln{\left(\Lambda^3 \rho(\mathbf{r},t)\right)} \Bigg)\Bigg].
\end{align}

In the ideal gas limit, where there are no inter-particle interactions i.e. $\mathcal{F}_{\text{exc}}[\rho]=0$, this equation reduces to a drift-diffusion equation where particles only diffuse and respond to an external potential. The DDFT equation in this standard form, with no velocity or inertial terms, describes purely overdamped dynamics where frictional forces dominate. It also relies on the adiabatic approximation which assumes that the non-equilibrium pair correlation function is equivalent to its equilibrium counterpart at the same one-body density. However, for systems \rev{where superadiabatic components such as rotational flows or memory effects are significant}, more advanced frameworks such as power functional theory~\cite{pft_bd, schmidt} and superadiabatic DDFT~\cite{tschopp} are required. It is also worth noting that since standard DDFT has the form of a continuity equation and conserves the particle number (neglecting the boundaries), it is only a canonical theory in this regard. In practice however it is common to still adopt the same grand-canonical free energy functional from cDFT under the adiabatic approximation. We follow this practice and adopt the equilibrium neural excess-free energy functional from Dijkman \textit{et al}.~\cite{dijdft}.

\section{Extension to open systems}\label{open}

Many practical systems are open to particle exchange with a reservoir---for instance, gas adsorption processes in porous materials---and these are often modeled using hybrid grand-canonical molecular dynamics simulations~\cite{gcmd, nemd}. We develop a novel extension of DDFT for such hybrid setups, allowing for particle insertion and removal through a reservoir coupling. This setting was studied before by Thiele \textit{et al.}~\cite{thiele}, where a source term was added to model the evaporation of a colloidal suspension into a vapor reservoir. We develop a similar extension, but from a principled Wasserstein-Fisher-Rao gradient flow perspective.

The 2-Wasserstein distance between two densities $\rho_0, \rho_1$ is defined as

\begin{equation}
\operatorname{W}_2^2(\rho_0, \rho_1) = \inf_{\rho \in \Gamma(\rho_0, \rho_1)} \int \rho(\mathbf{r}_1,\mathbf{r}_2)\Vert\mathbf{r}_1 - \mathbf{r}_2\Vert_2^2 d\mathbf{r}_1d\mathbf{r}_2,
\end{equation} where $\Gamma(\rho_0, \rho_1)$ is the space of all joint distributions with marginals $\rho_0$ and $\rho_1$. \rev{The infima gives the optimal coupling between the densities $\rho_0$, $\rho_1$ that minimizes their total expected Euclidean distance.} This can also be expressed in dynamical form as

\begin{equation}\label{wass_ot}
\operatorname{W}_2^2(\rho_0, \rho_1) = \inf_{v_t \in V(\rho_0, \rho_1), \rho_t} \int_0^1 \mathbb{E}_{\rho_t(\mathbf{r})} [\Vert v_t(\mathbf{r}) \Vert_2^2]dt,
\end{equation}
where $V(\rho_0, \rho_1)$ is the set of velocity fields $v_t$ such that $\left(\partial \rho_t/\partial t\right) = -\nabla_{\mathbf{r}}.(\rho_t(\mathbf{r})v_t(\mathbf{r}))$ with $\rho_{t=0} = \rho_0$ and $\rho_{t=1} = \rho_1$. \rev{In other words, the 2-Wasserstein distance is the optimal $(v_t, \rho_t)$ pair, satisfying the continuity equation, that minimizes the total expected kinetic energy.} From this dynamical formulation, which was first proposed by Benamou and Brenier~\cite{benamou2000computational}, it can be shown that the flow defined by DDFT [\cref{cddft}] is in fact the steepest descent on the free-energy functional $\mathcal{F}[\rho]$ in the space of densities under the 2-Wasserstein distance metric~\cite{jordan1998variational}.

The space of densities can be endowed with different metrics~\cite{ambrosio2008gradient} which correspond to different optimization dynamics. In particular, the Wasserstein Fisher-Rao distance~\cite{chizat2018interpolating, wasserstein} can be defined by extending ~\cref{wass_ot}, 

\begin{equation}
\operatorname{WFR}^2(\rho_0, \rho_1) = \inf_{v_t, \rho_t, g_t} \int_0^1 \mathbb{E}_{\rho_t(\mathbf{r})} [\Vert v_t(\mathbf{r}) \Vert_2^2 + \lambda g_t(\mathbf{r})^2]dt,
\end{equation} \rev{where $g_t$ is a growth term controlling the addition and removal of density, and $\lambda$ is a constant balancing the cost of the flow and growth terms. The} corresponding continuity equation is updated as $\left(\partial \rho_t/\partial t\right) = -\nabla_{\mathbf{r}}.(\rho_t (\mathbf{r}) v_t(\mathbf{r})) + g_t(\mathbf{r})\rho_t(\mathbf{r})$. \rev{Thus, the WFR distance allows for unbalanced transport where the total particle number can change.}
It can be shown that the flow that minimizes the free-energy functional $\mathcal{F}[\rho]$ under this Wasserstein Fisher-Rao metric is given by the following PDE (proof in~\cite{wasserstein}):

\begin{equation}
\begin{split}
\frac{\partial \rho_t(\mathbf{r})}{\partial t}
= &-\nabla_{\mathbf{r}} \cdot\left(\rho_t(\mathbf{r})\left(-\nabla_{\mathbf{r}} \frac{\delta \mathcal{F}\left[\rho_t\right]}{\delta \rho_t}(\mathbf{r})\right)\right) \\
& - \frac{1}{\lambda} \left(\frac{\delta \mathcal{F}\left[\rho_t\right]}{\delta \rho_t}(\mathbf{r}) - \mathbb{E}_{\rho_t(y)}\left[\frac{\delta \mathcal{F}\left[\rho_t\right]}{\delta \rho_t}(y)\right] \right) \rho_t(\mathbf{r}).
\end{split}
\end{equation}

\begin{figure*}
    \includegraphics[width=0.9\textwidth]{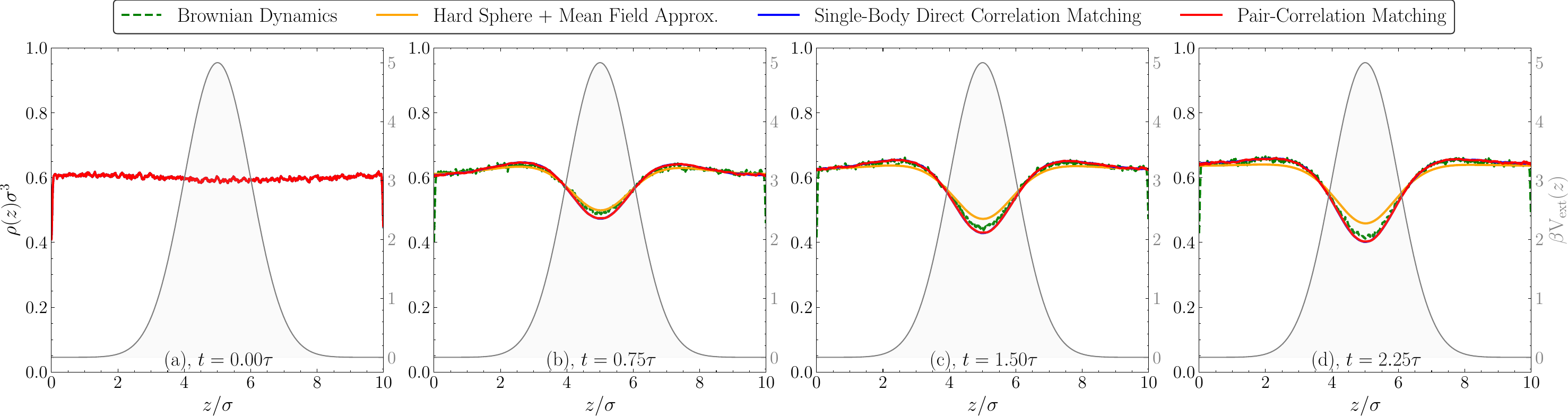}
    \caption{Time evolution of the one-body density profile $\rho(z, \tau)$ for a system of interacting $3$D LJ particles with planar geometry subjected to a repulsive Gaussian potential $V_\text{ext}(z)$ (grey curve, right axis). The plots compare two neural excess free-energy functionals and an analytical FMT functional, with a reference Brownian dynamics simulation. Initially ($t = 0\tau$), the system is at equilibrium with a flat density profile. Upon application of the external potential, the density profile evolves non-trivially, exhibiting a depletion near the potential maximum and eventually approaching a new equilibrium \rev{($t = 3\tau$)}. The neural DDFT approaches (red and blue) exhibit good agreement with Brownian dynamics throughout the temporal evolution, notably with the pair-correlation matching approach (red) not having seen any inhomogeneous densities during training.}
    \label{fig:snapshots}
\end{figure*}

While standard DDFT [\cref{cddft}] defines the change of density along a vector field $v_t(\mathbf{r})=\nabla_{\mathbf{r}} \left(\delta \mathcal{F}[\rho]/{\delta \rho_t}\right)$, this PDE also defines the addition and removal of density with a growth term that is proportional to the difference between the local chemical potential $\left({\delta \mathcal{F}\left[\rho_t\right]}/{\delta \rho_t}(\mathbf{r})\right)$ and the average chemical potential of the whole system. We replace this average potential with a target chemical potential $\mu_{\rev{\text{T}}}$ to model a grand canonical reservoir and get the following form, 

\begin{equation}\label{cddft_open}
\begin{split}
\frac{\partial \rho(\mathbf{r}, t)}{\partial t} =\,& 
\underbrace{\Gamma \nabla \cdot\Bigg[\rho(\mathbf{r}, t) \nabla\Bigg(
\frac{\delta \mathcal{F}[\rho]}{\delta \rho(\mathbf{r}, t)}\Bigg)\Bigg]}_\text{continuity equation} \\
&+ \underbrace{(1-\lambda(\mathbf{r})) \Gamma' \Bigg(\mu_{\rev{\text{T}}} -  
 \frac{\delta \mathcal{F}[\rho]}{\delta \rho(\mathbf{r}, t)} \Bigg)\rho(\mathbf{r}, t)}_\text{growth term},
\end{split}
\end{equation} where $\lambda(\mathbf{r})$ is a masking function such that $\lambda(\mathbf{r}) = 0$ for $\mathbf{r} = \mathbf{r}_i$, the locations where a particle bath is to be connected, and is equal to 1 elsewhere. $\Gamma'$ is the corresponding mobility coefficient that controls the rate of the non-conserved dynamics.

We know from cDFT that, at equilibrium, $\mu$ is given as, 
\begin{equation}
    \mu_{\rev{\text{eq}}} = V_{\text{ext}}\rev{(\mathbf{r})} + \frac{1}{\beta}  \ln{\left(\Lambda^3 \rho_{\text{eq}}(\mathbf{r})\right)} + \left.\frac{\delta \mathcal{F}_{\text{exc}}[\rho(\mathbf{r})]}{\delta \rho(\mathbf{r})}\right|_{\rho=\rho_\text{eq}}.
\end{equation} This relation \rev{shows} that at equilibrium the contribution from the external potential $V_{\text{ext}}\rev{(\mathbf{r})}$, ideal gas \rev{$\ln \left(\Lambda^3\rho_{\text{eq}}(\mathbf{r})\right)$}, and the functional derivative of the excess free energy add up at each position to $\mu_{\rev{\text{eq}}}$. Using this conservation relation we can write the growth-term PDE as,

\begin{equation}\label{gcpde}
\begin{split}
    \left.\frac{\partial \rho(\mathbf{r}, t)}{\partial t}\right|_{\rev{\text{gr.}}} = 
    -\rev{\Gamma'}\Bigg(&V_{\text{ext}}\rev{(\mathbf{r})} + \frac{1}{\beta} \ln{\left(\Lambda^3 \rho(\mathbf{r},t)\right)} \\
           &+ \frac{\delta \mathcal{F}_{\text{exc}}[\rho(\mathbf{r},t)]}{\delta \rho(\mathbf{r},t)} - \mu_{\rev{\text{T}}} \Bigg)\rho(\mathbf{r},t).
\end{split}
\end{equation}
Note that the steady-state solution of this PDE (considering the total $\mathcal{F}[\rho]$),
\begin{equation}
    \left\Vert \nabla \frac{\delta \mathcal{F}[\rho(\mathbf{r},t)]}{\delta \rho(\mathbf{r},t)} \right\Vert = 0 \iff \frac{\delta \mathcal{F}[\rho(\mathbf{r},t)]}{\delta \rho(\mathbf{r},t)} = \mu_{\rev{\text{T}}},
\end{equation}
is equivalent to the equilibrium density provided by the Euler-Lagrange equation in cDFT and hence it can also be interpreted as a continuous-time PDE alternative to Picard iterations.

\section{Implementation}

\begin{figure*}
    \includegraphics[width=0.9\textwidth]{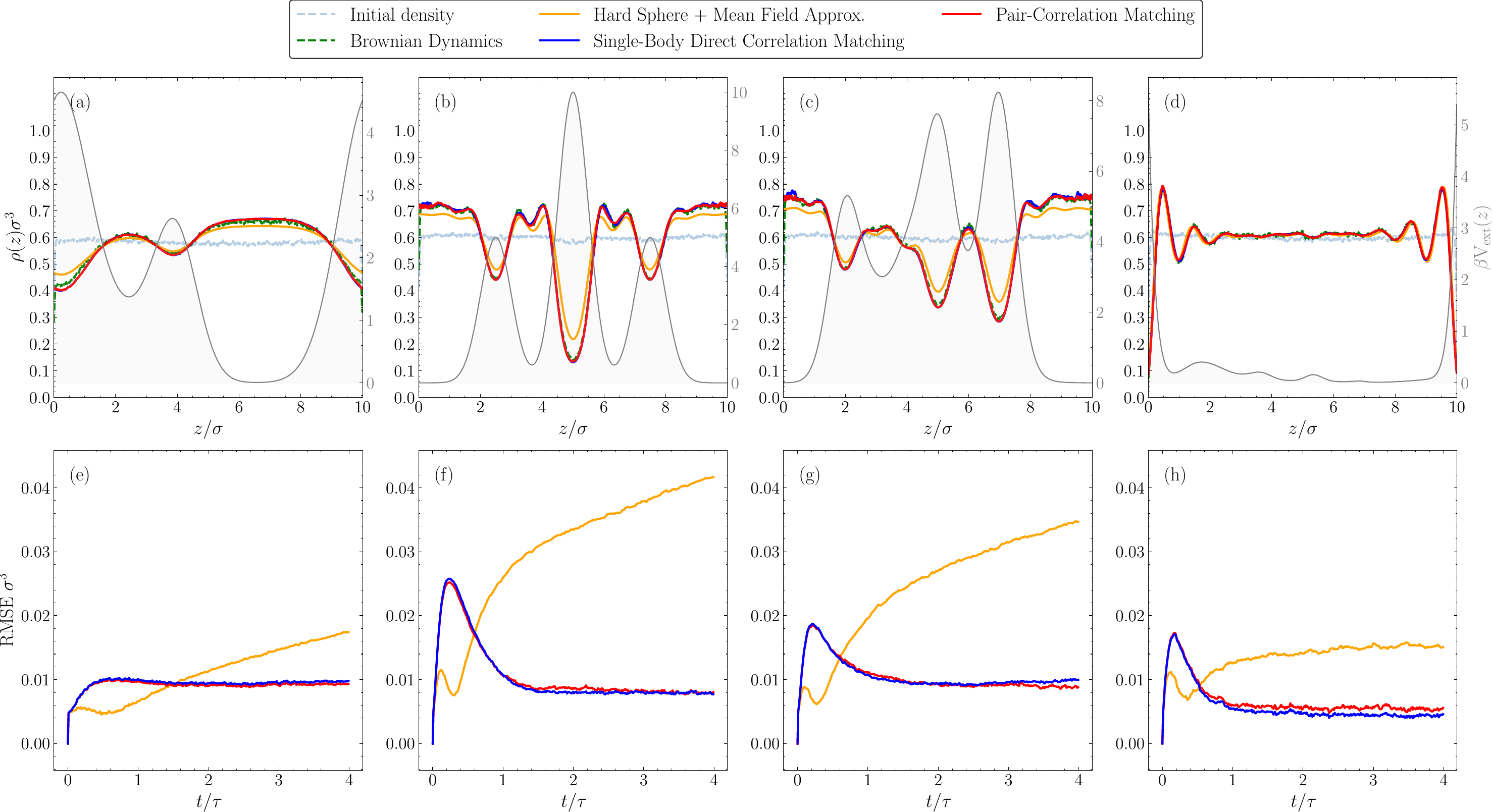}
    \caption{The final equilibrium density profiles and their RMSE over time evolution under complex external potentials (see SM~\cite{supplement} for the movie version). (a)-(d) A comparison of the final 1D equilibrium density profiles for a three-dimensional LJ system with planar geometry under different external potentials constructed from sums of repulsive Gaussian and steep well potentials. (e)-(h) show the corresponding RMSE deviation from the Brownian dynamics results over time evolution. The densities obtained with the neural functionals are significantly more accurate than those with the analytical functional, notably with the pair-correlation matching approach (red) not having seen any inhomogeneous densities during training.}
    \label{fig:grid}
\end{figure*}

We now describe the numerical implementation of our neural DDFT approach in relation to a Brownian dynamics (BD) reference simulation, where BD is chosen rather than (ballistic) molecular dynamics to be consistent with the overdamped nature of DDFT.

\subsection{Brownian dynamics} We use LAMMPS~\cite{LAMMPS} to simulate BD motion of Lennard-Jones (LJ) particles under repulsive external fields. The motion is specified by the following overdamped Langevin equation:

\begin{equation}
    d\mathbf{r} = \rev{\Gamma} \mathbf{F} \, dt + \sqrt{2k_B T \, \rev{\Gamma}} \, d\mathbf{W}_t,
\end{equation} where $\Gamma$ is the mobility coefficient, $\mathbf{F}$ is the force acting on the particle, and $d\mathbf{W}_t$ is a Wiener process.

We run BD simulations inside a closed $3$D cube box of edge length $10\sigma$ with periodic boundary conditions on all axes. We randomly initialize $600$ LJ particles (with a cut-off distance of $4\epsilon$) inside the box. $\epsilon$ and $\sigma$ are the LJ coefficients of well-depth and particle diameter respectively. The temperature of the system is kept at a constant $k_BT/\epsilon = 2$ and each timestep is $10^{-3} \tau$ where $\tau$ is the time scale $\rev{(\sqrt{{m\sigma^2}/{\epsilon}})}$.

We first run an energy minimization to adjust the random initial configuration to be in a local energy minimum with tolerance $10^{-4} \epsilon$, followed by $2000$ equilibration steps which results in a uniform density distribution. We then apply an external potential (mixture of repulsive Gaussian and steep well) on this system, using a PyLAMMPS extension~\cite{pylammps}, and run $4000$ collection steps ($4\tau$) with the thermodynamic output variables averaged over $10$ steps. The density is collected every $10$ steps, averaged over a bin size of $\sigma/100$ along the z-axis and across the xy plane, assuming planar geometry. We further average over $5000$ BD runs to obtain smooth density profiles.

\subsection{Dynamical density functional theory} We numerically integrate the DDFT equation [\cref{cddft_full}], with the neural excess free-energy functional from pair-correlation matching,
\begin{align}\label{cddft_theta}
    \frac{\partial \rho(\mathbf{r}, t)}{\partial t} &= \Gamma \nabla \cdot\Bigg[\rho(\mathbf{r}, t) \nabla \Bigg(\frac{\delta \mathcal{F}_{\theta, \text{exc}}[\rho]}{\delta \rho(\mathbf{r}, t)}
   + V_{\text{ext}}(\mathbf{r},t)  \nonumber \\
   &\qquad\qquad + \frac{1}{\beta} \ln{\left(\Lambda^3 \rho(\mathbf{r},t)\right)} \Bigg)\Bigg],
\end{align}to obtain time-dependent density profiles of $3$D Lennard-Jones (LJ) systems with planar geometry under an external potential $V_\text{ext}\rev{(\mathbf{r})}$ (\autoref{fig:block}).

We define a $1$D density grid of $10 \sigma$ length with a spacing of $\sigma/100$ for the solver, with periodic boundaries. We initialize the density grid with the same uniform initial density as the reference BD simulation and apply the same external potential. We simulate the PDE with a time-step of $10^{-4} \tau$ until $t=4\tau$, where $\tau = \sqrt{{m\sigma^2}/{\epsilon}}$ in reduced LJ units. The gradients are computed using finite-differences and time integrated using an Euler scheme, as implemented in the numerical Py-PDE Python package~\cite{py-pde}. A \rev{pseudospectral} solver can also be used for scenarios that require more numerical precision~\cite{pseudospectral}. The mobility coefficient $\Gamma$ is chosen to be the same as in BD. For computing ${\delta \mathcal{F}_{\theta, \text{exc}}[\rho(\mathbf{r}, t)]}/{\delta \rho(\mathbf{r}, t)}$, we use automatic differentiation over the excess neural free-energy functional.

\section{Experiments}\label{expts}

\begin{figure*}
    \includegraphics[width=0.9\textwidth]{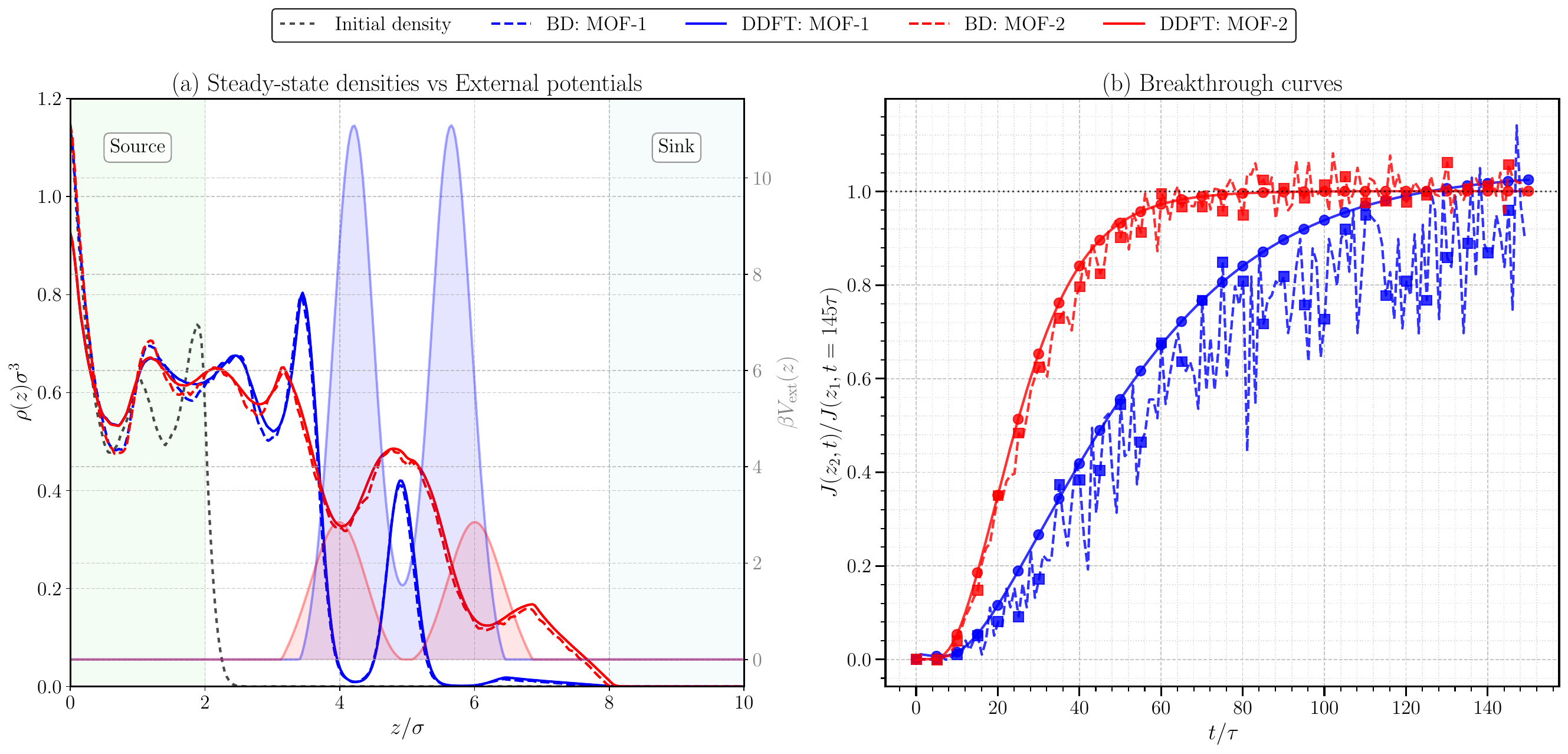}
    \caption{Breakthrough simulations using neural DDFT and Brownian dynamics (see SM~\cite{supplement} for the movie version). (a) LJ particles from the source move towards the sink under two different MOF-like repulsive external potentials. The dynamics of the density evolution and the final evolved density (visualized) obtained from DDFT are in close agreement with BD. (b) Time evolution of the output to input flux ratio, showing the approach towards steady-state or breakthrough condition. The particles reach breakthrough earlier under the MOF-2 external potential. The curves obtained from DDFT are in close agreement with BD, while being smoother and taking only $2\%$ of the compute time of BD.}
    \label{fig:break-curve}
\end{figure*}

We run two sets of ``experiments'' to evaluate the accuracy and efficiency of neural DDFT from pair-correlation matching with respect to BD, one on a closed system with planar geometry and conserved number of particles under various external potentials, and the other on an open system with a source and sink with varying number of particles, inspired by a gas adsorption process.

\subsection{Closed system} We use a 3D system of LJ particles with planar geometry, under external potentials that are mixtures of repulsive Gaussian and steep well potentials as in~\cite{dijdft}. In addition to the neural functional from pair-correlation matching (denoted by $\mathcal{F}_{\theta, \text{exc}}^{(2)}$), we also train a functional using single-body direct correlation matching (denoted by $\mathcal{F}_{\theta, \text{exc}}^{(1)}$) (see~\Cref{c1}). Both functionals use the same convolutional neural network with periodic and dilated convolutions, each with a kernel size of 3, a dilation of 2, and 6 layers. The layer dimensions are set to $N_\text{channels}$ = [16, 16, 32, 32, 64, 64], with average pooling with kernel size 2 applied after each layer. The models are trained for 5000 epochs, each taking less than an hour on an RTX 4070 GPU for convergence.

For training $\mathcal{F}_{\theta, \text{exc}}^{(2)}$, we simulated $1000$ \textit{uniform} densities and measured their radial distribution functions with ${\sigma}/{100}$ grid-size and chemical potentials in the range -4 to 0.5. $\mathcal{F}_{\theta, \text{exc}}^{(1)}$ is trained on $800$ non-uniform densities~\cite{dijdft} with ${\sigma}/{100}$ grid-size and chemical potentials in the range -4 to 0.5. The dataset of $\mathcal{F}_{\theta, \text{exc}}^{(2)}$ is larger, since no validation set had to be excluded from the train set, since it is validated on samples from the non-uniform dataset. We also evaluate against an analytical functional $\mathcal{F}^{\text{MF}}_{\text{exc}}$ which models the attractive contribution using mean-field approximation and the hard-sphere contribution using White-Bear II version of FMT~\cite{hs-fmt}, as implemented in PyDFTlj~\cite{Soares2023}.

Our experiments show that the neural free energy functionals $\mathcal{F}_{\theta, \text{exc}}^{(1)}$ and $\mathcal{F}_{\theta, \text{exc}}^{(2)}$ are both significantly better than the $\mathcal{F}^{\text{MF}}_{\text{exc}}$ functional in accuracy across time, as shown in~\autoref{fig:snapshots} and \autoref{fig:grid}. The convergence of $\mathcal{F}_{\theta, \text{exc}}^{(2)}$ is similar to $\mathcal{F}_{\theta, \text{exc}}^{(1)}$, except in Fig. 3(d) where the density near the wells is pushed above the maximum bulk value of $0.67\sigma^{-3}$ seen during training of $\mathcal{F}_{\theta, \text{exc}}^{(2)}$. The errors for all the DDFT approaches for time intervals $t<1\tau$ are high which we attribute to overly fast dynamics arising from the adiabatic approximation~\cite{tschopp}. More notably, the dynamical density profiles from DDFT are several orders of magnitude faster to compute with DDFT than with BD simulation: while running $5000$ trials of BD takes about a day on a single-core CPU, the DDFT approach on the same computer takes less than 6 minutes (see~\Cref{runtime} for runtime comparisons).

\subsection{Open system}\label{breakthrough} To evaluate the pair-correlation matching functional $\mathcal{F}_{\theta, \text{exc}}^{(2)}$ on an open system, we construct a system modeled after ``breakthrough'' experiments used in gas adsorption studies~\cite{mof-breakthrough, ruptura}. In these studies, typically a column filled with adsorbent material [e.g. nanoporous metal-organic framework (MOF)] is exposed to a gas stream (e.g. CO\textsubscript{2}), where the breakthrough of the gas is detected at the outlet to estimate the adsorption capacity and kinetics of the material. We adopt such a setup by extending the closed Lennard-Jones system from the previous experiment with two particle baths, one at a high $\rev{\beta} \mu=0.5$ in $z\in[0,2]\sigma$ to simulate an LJ particle source and the other at extremely low $\rev{\beta} \mu=-100$ in $z\in[8,10] \sigma$ to simulate an LJ particle sink. The difference in $\mu$ causes particles to flow from the source to the sink through the region  $z\in[2,8] \sigma$, in which we apply, in addition, a planar potential to mimic two MOF planes from the IRMOF series~\cite{irmof} (see~\Cref{mof-potential}).
In the reference BD simulation, we extend the closed system to an open system by attaching two Grand-Canonical Monte-Carlo (GCMC) reservoirs as source and sink. The source reservoir is from $z=[0,2]\sigma$ with $\rev{\beta} \mu=0.5$ and the sink is from $z=[8,10]\sigma$ with $\rev{\beta} \mu=-100$. Ten trials of GCMC insertion/deletion steps, with no translation/rotation, are attempted in LAMMPS (hybrid) between every BD update. Before starting the breakthrough simulation, we first randomly initialize LJ particles in the source region and confine them using a potential barrier and equilibrate the system for $20000$ steps (each timestep is $10^{-3}\tau$). Then the barrier is removed and the MOF-like potential is applied on the system and $100000$ collection steps are performed as the particles flow towards the sink. The planar density is sampled every 100 steps, over a bin size of ${\sigma}/{32}$ and is averaged over $500$ BD runs.

We simulate this same system using our open DDFT equation [\cref{cddft_open}] for a duration of $150\tau$, with a timestep of $10^{-3} \tau$, grid spacing of ${\sigma}/{32}$, and $\Gamma' = 1$. To obtain breakthrough curves we compute the time-dependence of the ratio $J(z_2,t)/J(z_1,T)$ of the particle fluxes $J(z_2,t)$ close to the output at $z_2=7.5\sigma$ and $J(z_1,T)$ close to the input at $z_1=2.5\sigma$ at late time $T=145\tau$ where steady-state is (almost) reached. The particle flux from DDFT is computed as $J(z,t) = -\Gamma \rho(z,t) \nabla \left({\delta \mathcal{F}[\rho(z)]}/{\delta \rho(z)}\right)$, where $\mathcal{F}_{\theta, \text{exc}}^{(2)}$ is used for the excess component. We also apply a Savitsky-Golay filter~\cite{sg-filter} over the spatial gradient in the flux computation to smooth out any noise arising from numerical errors. The BD flux is computed as the net number of particles crossing the planes at $z_1$ and $z_2$ per unit time per unit area, averaged over $500$ BD runs.

As observed from the breakthrough curves in~\autoref{fig:break-curve}, the dynamics of both the BD and DDFT simulations agree closely and reach breakthrough at similar times. The potential from MOF-2 (red), which corresponds to a MOF with larger pore size, presents a lower particle barrier giving rise to a faster evolution to the steady state, which is captured by both simulations. But the DDFT simulation is much more efficient, taking only $2\%$ of the compute time of the BD simulation (see~\Cref{runtime} for runtime comparisons). Note that since the GCMC moves are run in between each BD update while the open DDFT equation updates are simultaneous, we do not expect exact dynamical agreement though this can be alleviated to some extent with the choice of $\Gamma'$ in the growth term.
\section{Conclusion}

To summarize, we have presented results that the neural excess free-energy functional trained via pair-correlation matching can be directly applied to DDFT without retraining, with high accuracy relative to an analytical free-energy approximation and with high efficiency relative to BD simulations. We show this for both conserved dynamics, as well as for non-conserved dynamics via a novel gradient flow based extension which we evaluate on a system inspired by breakthrough adsorption simulations. \rev{Pair-correlation matching thus offers a practical route to predicting non-equilibrium dynamics by training on homogeneous equilibrium density data, which is significantly less expensive to sample in higher dimensions than inhomogeneous equilibrium density data~\cite{dijdft}, inhomogeneous steady-state density data~\cite{zimmermann2024neural} or out-of-equilibrium dynamical density data}.

Future work could focus on extending this method to incorporate superadiabatic forces~\cite{heras}. \rev{In this work, we relied on the adiabatic approximation of DDFT, i.e. the non-equilibrium internal force field can be approximated as a conservative force given by the gradient of the equilibrium single body direct correlation function, to drive the non-equilibrium dynamics. Non-conservative forces can be important, especially for shear flows, which requires learning a correction term to this equilibrium functional. However the main challenge in training such a correction term is that it would require training on dynamical trajectories from BD, which can be very expensive to simulate relative to equilibrium densities from GCMC. To address this, Zimmermann \textit{et al}.~\cite{zimmermann2024neural} recently proposed training a neural force functional on \textit{steady-state} dynamical data, which are relatively easier to sample than non-equilibrium dynamical data, showing promising results in extrapolating to out-of-steady-state non-equilibrium dynamics with non-conservative forces.} Other future work includes extensions to systems with full $3$D geometry, anisotropic particles, and mixtures towards practical applications in modeling many-body non-equilibrium systems.

\includecomment{This paragraph does not yet feel right to me yet. Maybe it needs one sentence explaining that steady state systems are easier to sample than non-equilibrium states. Also, maybe there needs to be a short explanation on that this method still has difficulty on out-of-equilibrium states far from steady-state. Maybe there needs to be one sentence like "in this sense, this paper demonstrates a more extreme version, going one step further by extrapolating from bulk to non-equilibrium instead of steady-state to equilibrium.} 

\section*{Acknowledgements}We would like to thank David Dubbeldam and Christian Koke for useful discussions. J.D. and B.E. would like to thank the University of Amsterdam Data Science Centre for financial support. K.R. acknowledges his affiliation with the ELLIS (European Laboratory for Learning and Intelligent Systems) PhD program.

\section*{Data availability}The source code for BD and DDFT simulations, along with the model weights and resulting trajectories are made available at~\cite{zenodo}.

\appendix
\section{Classical Density Functional Theory}\label{cdft_theory}

Classical density functional theory (cDFT)~\cite{evans} proves the existence of a grand potential functional $\Omega[\rho]$ of a one-body density profile $\rho(\mathbf{r})$, with the condition that the equilibrium density profile $\rho_\text{eq}(\mathbf{r})$ minimizes $\Omega[\rho]$ and is equal to the equilibrium grand potential $\Omega_0$. From $\Omega_0$, all the thermodynamic properties of the system of interest can be subsequently derived.

\begin{equation}\label{elag}
\left.\frac{\partial \Omega(\mu,[\rho], T)}{\partial \rho(\mathbf{r})}\right|_{\rho=\rho_{\text{eq}}}=0, \quad \Omega([\rho_{\text{eq}}]) = \Omega_0.
\end{equation}
The grand potential functional $\Omega[\rho]$ can be written as,
\begin{equation}
\Omega(\mu,[\rho], T)=\mathcal{F}([\rho], T)- \int \mathrm{d}^3 r \rho(\mathbf{r})(\mu - V_{\text{ext}}(\mathbf{r})),
\end{equation}
where $\mu$ is the chemical potential, $T$ is the temperature, and $\mathcal{F}[\rho]$ is the intrinsic Helmholtz free-energy functional which can be split into three parts,

\begin{equation}\label{fenergy}
    \mathcal{F}([\rho], T) = \mathcal{F}_{\text{id}}([\rho], T) + \mathcal{F}_{\text{ext}}([\rho], T) + \mathcal{F}_{\text{exc}}([\rho], T),
\end{equation}
where $\mathcal{F}_{\text{id}}([\rho], T) = \frac{1}{\beta}\int d^3\mathbf{r}\rho(\mathbf{r})(\ln(\Lambda^3 \rho(\mathbf{r})) - 1)$, $\beta = \left(1/{k_BT}\right)$ and $\Lambda$ is the thermal wavelength, and $\mathcal{F}_{\text{ext}}([\rho]) = \int d^3\mathbf{r} \rho(\mathbf{r}) V_{\text{ext}}(\mathbf{r})$.

The challenge is to approximate $\mathcal{F}_\text{exc}[\rho]$ which represents the interparticle interactions as a functional of the non-local density. Once such an $\mathcal{F}[\rho]$ is (approximately) obtained, the equilibrium density $\rho_{\text{eq}}(\rho)$ from \autoref{elag} can be formulated as the following Euler-Lagrange equation,

\begin{equation}\label{cdft}
\rho_{\text{eq}}(\mathbf{r})=\frac{1}{\Lambda^3} \exp \left(\beta \mu -\left.\beta \frac{\delta \mathcal{F}^{e x c}[\rho]}{\delta \rho(\mathbf{r})}\right|_{\rho=\rho_\text{eq}}-\beta V_{\text{ext}}(\mathbf{r})\right).
\end{equation}
This self-consistency relation can be solved by means of a recursive Picard iteration scheme to compute the equilibrium density starting from any arbitrary initial density. It is also worth noting that cDFT can be interpreted as a grand-canonical framework in which the system is in thermal and diffusive equilibrium with a particle bath at chemical potential $\mu$ and temperature $T$.

\begin{figure}
    \centering
    \includegraphics[width=\linewidth]{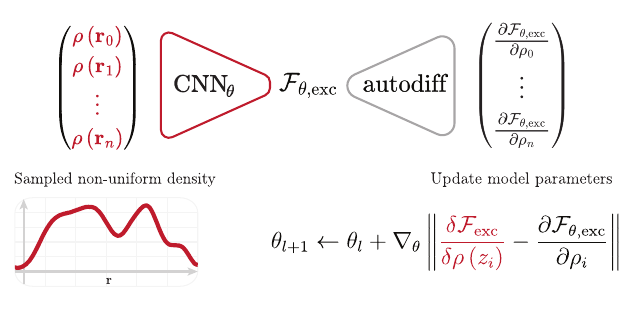}
    \caption{An overview of single-body direct correlation matching.}
    \label{sbc-fig}
\end{figure}

\subsection{Single-body direct correlation matching}\label{c1}

In addition to pair-correlation matching~\cite{dijdft}, we can also train a neural free-energy functional ($\mathcal{F}_{\theta, \text{exc}}^{(1)}$) using single-body direct correlation matching (\autoref{sbc-fig}). This is done by minimizing the error between the target ${\delta \mathcal{F}_\text{exc}}/{\delta \rho(z_i)}$ and the estimate $(1/{\Delta z}){\partial \mathcal{F}_{\theta, \text{exc}}^{(1)}}/{\partial \rho_i}$. This functional derivative of the excess free-energy is the one-body direct correlation function, and computing the target samples requires simulating inhomogeneous equilibrium densities under a range of external potentials.

\section{Dynamical Density Functional Theory}\label{ddft-derivation}

We provide here a phenomenological derivation for classical dynamical density functional theory (DDFT) starting from cDFT, originally presented by Evans~\cite{evans}. DDFT can also alternatively be derived from Langevin equations~\cite{marconi}, Smoluchowski equations~\cite{archer2004dynamical}, as well as from a projection operator formalism~\cite{yoshimori2005microscopic}. Consider the continuity equation for conserved $\rho$,

\begin{equation}
\frac{\partial \rho(\mathbf{r}, t)}{\partial t} + \nabla \cdot \mathbf{J}(\mathbf{r}, t) = 0.\\
\end{equation}
Considering the flux $\mathbf{J}(\mathbf{r},t) = \rho(\mathbf{r}, t) \mathbf{v}(\mathbf{r}, t)$,

\begin{equation}
\frac{\partial \rho(\mathbf{r}, t)}{\partial t} + \nabla \cdot \left(\rho(\mathbf{r}, t) \mathbf{v}(\mathbf{r},t)\right) = 0,\\
\end{equation}
where the average velocity $\mathbf{v}(\mathbf{r},t) = - \Gamma \nabla \mu(\mathbf{r}, t)$ with the mobility coefficient $\Gamma$. Expanding this we get,

\begin{equation}
\frac{\partial \rho(\mathbf{r}, t)}{\partial t} - \Gamma \nabla \cdot \left(\rho(\mathbf{r},t) \nabla \mu(\mathbf{r},t)\right) = 0.\\
\end{equation}
From cDFT, we know at equilibrium $\mu = {\delta \mathcal{F}[\rho(\mathbf{r})]}/{\delta \rho(\mathbf{r})}$. This equilibrium density functional can also be applied to DDFT close to equilibrium. This leads to,

\begin{equation}\label{ddft_eq}
    \frac{\partial \rho(\mathbf{r}, t)}{\partial t} = \Gamma \nabla \cdot\left[\rho(\mathbf{r}, t) \nabla\left(\frac{\delta \mathcal{F}[\rho(\mathbf{r}, t)]}{\delta \rho(\mathbf{r}, t)}\right)\right].
\end{equation}
Re-arranging the terms and expanding the free-energy functional we obtain,

\begin{align}
    \frac{\partial \rho(\mathbf{r}, t)}{\partial t} &= \Gamma \nabla \cdot\Bigg[\rho(\mathbf{r}, t) \nabla \Bigg(\frac{\delta \mathcal{F}_\text{exc}[\rho]}{\delta \rho(\mathbf{r}, t)}
   + V_{\text{ext}}(\mathbf{r},t)  \nonumber \\
   &\qquad\qquad + \frac{1}{\beta} \ln{\left(\Lambda^3 \rho(\mathbf{r},t)\right)} \Bigg)\Bigg].
\end{align}

\section{MOF-like planar potentials}\label{mof-potential}

Here we describe our procedure for obtaining the $1$D external potentials from metal-organic frameworks (MOF) for the breakthrough simulation experiment in the main paper. 

\begin{figure}
    \centering
    \includegraphics[width=0.9\linewidth]{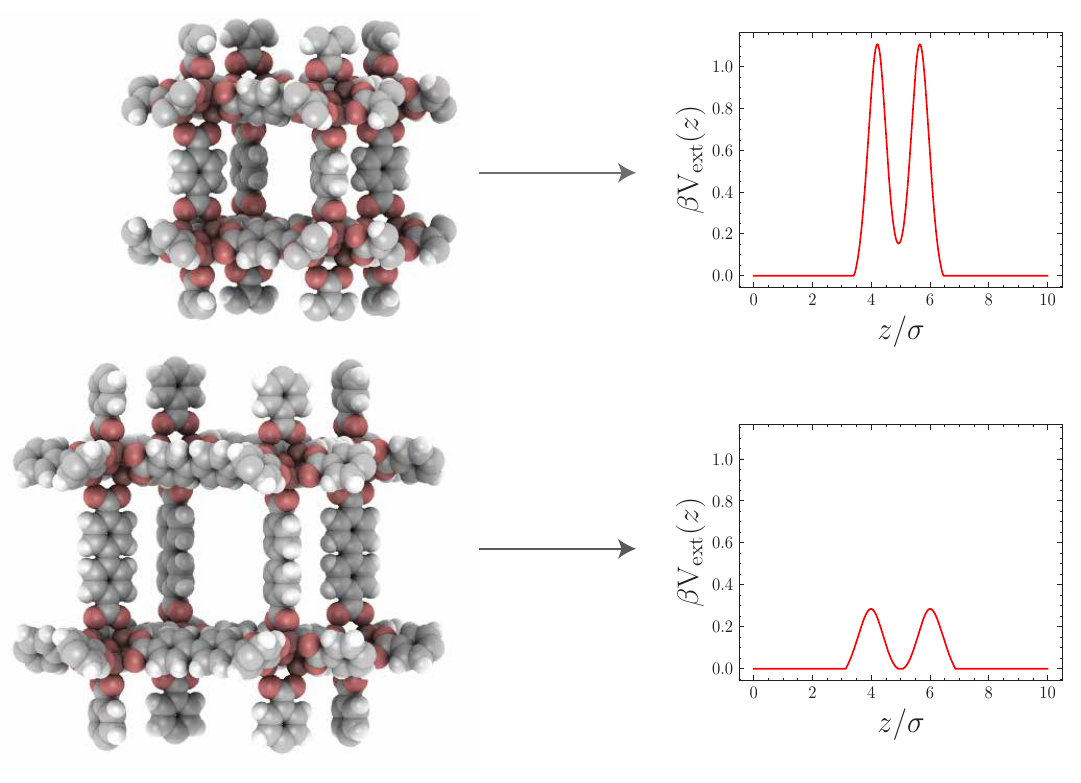}
    \caption{IRMOF-1 (top) and IRMOF-10 (bottom) and their corresponding effective 1D potentials computed along the pore axes.}
    \label{fig:irmof-projects}
\end{figure}

We use the IRMOF-1 and IRMOF-10 structures (\autoref{fig:irmof-projects}) from the IRMOF series of crystal structures~\cite{irmof}, which have the same topology but differing pore sizes. To convert their 3D structures into effective 1D external potentials suitable for DDFT, we employed a simulation-based force sampling approach using LAMMPS~\cite{LAMMPS}. Starting from CIF files of the MOFs, we generated compatible data files for LAMMPS with appropriate force field assignments and topology using lammps-interface~\cite{lammps-interface}. The pore axis of each MOF was aligned along the z-direction, and a single LJ-type guest particle was introduced and displaced incrementally along its pore axis and the total force acting on it was recorded at each step. This produced a raw force profile as a function of z-position which was then converted to reduced LJ units and numerically integrated (in reverse) to generate a dimensionless potential profile. The resulting potentials were clipped and centered to remove artifacts at the boundaries and padded with zeros to enforce a flat, finite domain. The final 1D potentials, expressed as functions interpolated over a grid, serve as external fields $V_\text{ext}\rev{(\mathbf{r})}$ that approximate the confinement environment of the original MOF pores.

\section{Runtime comparisons}\label{runtime}
  
The table below is a comparison of \textit{approximate} runtimes between BD simulation and our DDFT implementation for open and closed systems across two different grid sizes. While each DDFT simulation directly produces smooth density profiles, each BD simulation needs to be ensemble averaged (over 5000 and 500 runs for closed and open, respectively), to obtain comparable density profiles. Both simulations were run on a single-core of Intel(R) Xeon(R) W-2123 CPU @ 3.60GHz or comparable CPU. Note that DDFT uses $\Delta t = 10^{-3}\tau$ for grid spacing ${\sigma}/{32}$ and $\Delta t = 10^{-4}\tau$ for grid spacing ${\sigma}/{100}$, while BD uses $\Delta t = 10^{-3}\tau$ for both.
\vspace{1em}

\begin{table}[H]
    \centering
    \ra{1.3}
    \setlength{\tabcolsep}{8pt}
    \begin{tabular}{lccc}
    \toprule
    {System} & {Grid} & {BD} & {DDFT} \\
    \midrule
    Closed [Eq. (M1), & ${\sigma}/{32}$ & 34 h & 25 s \\
    until $t=4\tau$]  & ${\sigma}/{100}$ & 34 h & 5 min \\
    \midrule
    Open [Eq. (M2),  & ${\sigma}/{32}$  & 48 h & 40 min  \\
    until $t=150\tau$] & ${\sigma}/{100}$ & 48 h & 8 h  \\
    \bottomrule
    \end{tabular}
    \label{tab:bd_vs_ddft}
\end{table}





\bibliography{references}

\end{document}